\documentclass[10pt,british,english,aps,superscriptaddress,preprintnumbers,amsmath,nofootinbib,amssymb,altaffilletter,twocolumn]{revtex4}
\usepackage[T1]{fontenc}
\usepackage[latin9]{inputenc}
\usepackage{xcolor}
\usepackage{babel}
\usepackage{float}
\usepackage{amsmath}
\usepackage{amssymb}
\usepackage{graphicx}
\usepackage{esint}
\usepackage[unicode=true,pdfusetitle,
 bookmarks=true,bookmarksnumbered=false,bookmarksopen=false,
 breaklinks=false,pdfborder={0 0 1},backref=false,colorlinks=false]
 {hyperref}

\makeatletter
\@ifundefined{textcolor}{}
{%
 \definecolor{BLACK}{gray}{0}
 \definecolor{WHITE}{gray}{1}
 \definecolor{RED}{rgb}{1,0,0}
 \definecolor{GREEN}{rgb}{0,1,0}
 \definecolor{BLUE}{rgb}{0,0,1}
 \definecolor{CYAN}{cmyk}{1,0,0,0}
 \definecolor{MAGENTA}{cmyk}{0,1,0,0}
 \definecolor{YELLOW}{cmyk}{0,0,1,0}
}

\usepackage{babel}
\usepackage{babel}
\usepackage{babel}
\usepackage{babel}

\usepackage{babel}
\usepackage{breakurl}

\newcommand{\ba}{\begin{aligned}}
	\newcommand{\ea}{\end{aligned}}

\newcommand{\cF}{{\cal F}}

\newcommand{\LF}{\left(}
\newcommand{\RF}{\right)}

\@ifundefined{textcolor}{}{%
 \definecolor{BLACK}{gray}{0}
 \definecolor{WHITE}{gray}{1}
 \definecolor{RED}{rgb}{1,0,0}
 \definecolor{GREEN}{rgb}{0,1,0}
 \definecolor{BLUE}{rgb}{0,0,1}
 \definecolor{CYAN}{cmyk}{1,0,0,0}
 \definecolor{MAGENTA}{cmyk}{0,1,0,0}
 \definecolor{YELLOW}{cmyk}{0,0,1,0}
}

\usepackage{babel}
\usepackage{epsfig}\usepackage{mathrsfs}\usepackage{euscript}

\renewcommand{\[}{\begin{equation}}
\renewcommand{\]}{\end{equation}}
\def\beq{\begin{equation}}
\def\eeq{\end{equation}}
\newcommand{\be}{\begin{eqnarray}}
\newcommand{\ee}{\end{eqnarray}}

\renewcommand{\texttt}{{}}

\def\bs{\begin{subequations}}
\def\es{\end{subequations}}

\def\cF{\mathcal{F}}

\def\Fc{\mathcal{F}}

\def\Kc{\mathcal{K}}
\def\Lc{\mathcal{L}}

\def\Zc{\mathcal{Z}}

\newcommand{\tia}[1]{}

\newcommand{\bea}{\begin{eqnarray}}
\newcommand{\eea}{\end{eqnarray}}
\newcommand{\beas}{\begin{eqnarray*}}
\newcommand{\eeas}{\end{eqnarray*}}
\newcommand{\bal}{\begin{aligned}}
\newcommand{\eal}{\end{aligned}}

\def\({\left(}
\def\){\right)}

\newcommand{\pd}{\partial}

\makeatother

\begin{document}

\title{Stable, non-singular bouncing universe with only a scalar mode}

\author{K. Sravan Kumar}
\email{sravan.korumilli@rug.nl}

\selectlanguage{british}%


\selectlanguage{english}%

\author{Shubham Maheshwari}
\email{s.maheshwari@rug.nl}

\selectlanguage{british}%

\selectlanguage{english}%

\author{Anupam Mazumdar}
\email{anupam.mazumdar@rug.nl}

\selectlanguage{british}%


\author{Jun Peng}
\email{jun.peng@rug.nl}

\selectlanguage{british}%

\address{Van Swinderen Institute, University of Groningen, 9747 AG Groningen, The Netherlands }
\selectlanguage{english}%

\begin{abstract}
In this paper, we study a class of higher derivative, non-local gravity which admits homogeneous and isotropic non-singular, bouncing universes in the absence of matter. At the linearized level, the theory propagates only a scalar degree of freedom, and no vector or tensor modes. The scalar can be made free from perturbative ghost instabilities, and has oscillatory and bounded evolution across the bounce.
\end{abstract}
\maketitle


Bouncing cosmologies have been an attractive paradigm for resolving the Big Bang singularity, which is inevitable in General Relativity (GR) under reasonable assumptions~\cite{Brandenberger:2016vhg}. They have been studied in various approaches to quantum gravity like string theory \cite{Gasperini:2002bn,Khoury:2001wf} and loop quantum gravity \cite{Ashtekar:2008zu}.
Bouncing solutions are typically found by modifying the matter sector by adding fields which violate the null energy condition (NEC), and/or by modifying GR, for instance, by adding higher derivative terms to the Einstein-Hilbert (EH) action \cite{Buchbinder:2007ad,Lin:2010pf,Brandenberger:2016vhg,Battefeld:2004cd,Gasperini:2003pb,Cartier:2001is,Bojowald:2007hv}. NEC violation and higher derivative actions generally imply sicknesses in the form of ghost (fields with wrong-sign kinetic terms) and gradient instabilities. Another problem with bouncing cosmologies is the growth of vector (metric and matter) modes with time during the contraction phase, thereby invalidating perturbation theory \cite{Battefeld:2004cd}. In some bouncing models, scalar modes for a certain range of wavelengths can also grow \cite{Xue:2013bva}. Of central importance is then the question of finding a healthy, non-singular bouncing scenario with bounded amplitudes of perturbations.

In the context of higher derivative modified gravity, quadratic scalar curvature ($R+R^2$) is an interesting model to consider when hunting for non-singular solutions \cite{Starobinsky:1980te,Starobinsky:1987zz}. After adding a cosmological constant to $R+R^2$ theory, one can indeed obtain bouncing solutions, but which nevertheless are plagued by the presence of negative radiation energy density $\rho <0$. Extending the action by adding an arbitrary number of higher derivative terms typically leads to ghost degrees of freedom around, for instance, flat space and (A)dS. On the other hand, an infinite covariant derivative, non-local extension can potentially resolve the problem of ghosts, by choosing a suitable function that captures all order derivative terms in the action. In this non-local quadratic scalar curvature  model, bouncing solutions were found with $\rho>0$, around spatially flat, homogeneous and isotropic backgrounds \cite{Biswas:2005qr,Biswas:2010zk,Biswas:2011qe,Biswas:2011ar,Koshelev:2012qn,Koshelev:2013lfm}. The stability of scalar perturbations around these solutions in the presence of radiation was studied in \cite{Biswas:2010zk,Biswas:2012bp}. One can deduce that the presence of radiation implies growth of vector perturbations during the contraction phase \cite{Craps:2014wga}.

So far, infinite derivative gravity theories have been constructed to be ghost-free around flat space~\cite{Biswas:2011ar,Biswas:2013kla,Tomboulis:1997gg,Modesto:2011kw} and (A)dS~\cite{Biswas:2016etb,Biswas:2016egy}. In an arbitrary background, inevitable mixing of scalar-vector-tensor (SVT) modes makes the task of removing ghosts challenging, but one can still remove them around specific non-maximally symmetric backgrounds \cite{SravanKumar:2019eqt}.

In this paper, we find a class of non-local $R+R{\cal F}R$ theory with a cosmological constant which is ghost-free around homogeneous and isotropic bouncing cosmological backgrounds. ${\cal F}$ is the non-local term (discussed below) essential for avoiding ghosts. To the best of our knowledge, we show for the first time that a class of higher derivative gravity can give stable, non-singular bouncing scenarios, in the absence of matter, and without any vector or tensor modes at the linearized level around a time dependent background. Only one scalar mode survives, which can be made ghost-free, and exhibits bounded, oscillatory evolution across the bounce. The metric signature is $(-+++)$, overbars on quantities like $\bar{R}$ indicate their background value, and $\hbar = c = 1$.



Non-local generalization of $R$ + $R^{2}$ by adding an infinite series of covariant derivatives reads \cite{Biswas:2005qr}
\begin{equation} \label{action1}
S = \int d^{4}x\sqrt{-g}\left[ \frac{M_p^{2}}{2} R + R \cF (\square) R  -  \Lambda  \right]
\end{equation}
where $\Lambda$ is the cosmological constant, the form factor $\Fc(\square)$ is an arbitrary analytic\footnote{Analyticity is needed to recover GR at low energies} function of d'Alembertian, $\square=g_{\mu\nu}\nabla^{\mu}\nabla^{\nu}$, containing all orders of $\square$. It has the power series expansion $\Fc(\square) =\sum_{n=0}^{\infty}f_{n}\square_s^n$, where $f_{n}$ are dimensionless coefficients, dimensionless $\square_{s} = \square/M_s^2$, and $M_s (<M_{p}\sim 10^{18}$ GeV) is a new high energy scale of non-locality below the Planck scale $M_p$. The equations of motion (EOM) of Eq.(\ref{action1}) are \cite{Biswas:2013cha}
\begin{equation}
\begin{aligned}
E_{\ \nu}^{\mu}\equiv & -\left[M_{p}^{2}+4\Fc(\square)R\right]G_{\ \nu}^{\mu}- R\Fc(\square)R\delta_{\ \nu}^{\mu} \\
&+4\left(\nabla^{\mu}\partial_{\nu}-\delta_{\ \nu}^{\mu}\square\right)\Fc(\square)R
 +2\mathcal{K}_{\ \nu}^{\mu}\\
  &-\delta_{\ \nu}^{\mu} (\mathcal{K}_{\ \sigma}^{\sigma}+\tilde{\mathcal{K}})-\Lambda \delta^\mu_{\ \nu} = 0
\ea
\label{EoM}
\end{equation}
where
$
\mathcal{K}_{\ \nu}^{\mu}  =   \frac{1}{M_{s}^{2}}\sum_{n=1}^{\infty}f_{n}\sum_{l=0}^{n-1} ( \partial^{\mu}\square_{s}^{l}R) \ (\partial_{\nu}\square_s^{n-l-1}R)
$
and
$
\mathcal{\tilde{K}} =  \sum_{n=1}^{\infty}f_{n}\sum_{l=0}^{n-1} ( \square_{s}^{l} R) \ (\square_{s}^{n-l}R).
$
The trace EOM is
\begin{equation}
\left(M_p^2 -12\square\Fc(  \square )\right) R- 2\Kc^\mu_{\ \mu}-4\tilde{\Kc}-4\Lambda= 0.
\label{trace}
\end{equation}
The structure of the form factor $\Fc(\square)$ is determined by finding the second order action around a fixed background and demanding that all perturbations be ghost-free. Note that most of the upcoming computations are carried out covariantly.
Solving the EOM Eq.(\ref{EoM}) in full generality is non-trivial but the following ansatz for the background Ricci scalar helps in making headway \cite{Biswas:2005qr,Biswas:2010zk,Biswas:2012bp}
\be
\bar{\square} \bar{R}= r_1 \bar{R} + r_2
\label{ansatz}
\ee
where $r_{1}$ and $r_{2}$ are dimensionful constant parameters. The ansatz Eq.(\ref{ansatz}) is essentially the trace EOM of local $R + R^{2}$ theory with a cosmological constant given by $\Lc = \frac{M_p^{2}}{2} R + f_{0} R^{2}-  \Lambda$ \cite{Biswas:2005qr}. The ansatz is solved by bouncing backgrounds with cosine hyperbolic and exponential type scale factors, as we will see soon. Straightforwardly, we then obtain
\begin{equation}
\bar{\square}^n \bar{R}= r_1^n \left(\bar{R}+\frac{r_2}{r_1} \right) \implies \Fc (\bar{\square}) \bar{R} = \Fc_1  \bar{R}+\Fc_2
\end{equation}
where
\begin{equation}
\Fc_1 = \Fc (r_{1})\qquad \text{and} \qquad
\Fc_2  = \frac{r_2}{r_1} ( \Fc_{1} - f_0 )
\label{F2}
\end{equation}
are constants. Substituting the ansatz Eq.\eqref{ansatz} in the EOM Eq.\eqref{EoM} and trace EOM Eq.\eqref{trace} gives us the following unique conditions on the form factor $\Fc(\bar{\square})$
\begin{equation}
\Fc_{1} = \Fc' (r_{1}) = 0,\quad \Fc_2= -\frac{M_p^2}{4} ,\quad  \Lambda = -\frac{M_p^2}{4} \frac{r_2}{r_1}
\label{condi}
\end{equation}
where $\cF'(r_{1})$ denotes the derivative of $\cF(\bar{\square})$ with respect to $\bar{\square}_{s}$, evaluated at $\bar{\square}=r_{1}$. From Eqs.~(\ref{F2}) and (\ref{condi}), we can deduce that 
\begin{equation}
f_0 <0\quad \text{for} \quad \Lambda>0\,,\qquad \ f_0>0\quad \text{for}\quad  \Lambda<0\,.
\label{pcc}
\end{equation}
Note that in Ref.\cite{Biswas:2010zk,Biswas:2012bp}, bouncing solutions and linear perturbations were studied in the presence of fluid radiation, which corresponds to the case when $\Fc_{1}<0$. In this paper, we emphasize that a bounce can be achieved in the theory satisfying the conditions in Eq.\eqref{condi}, without requiring any radiation/matter.
%
%
%
%
%
%

We now study perturbations and their stability around backgrounds satisfying the ansatz Eq.(\ref{ansatz}) in two, equivalent ways - both at the linearized level of EOM, and at the quadratic level of the action. Linearizing the full EOM Eq.(\ref{EoM}) and trace EOM Eq.(\ref{trace}) around the background given by Eq.(\ref{ansatz}), and substituting the conditions in Eq.\eqref{condi}, we obtain, respectively
\begin{widetext}
\begin{equation}
\begin{aligned}
\delta E^\mu_{\ \nu}  = & \Bigg[ \Big(4\bar{G}^\mu_{\ \nu}  + 2\delta^\mu_{\ \nu}\bar{ R}  - 4 (  \bar{\nabla}^\mu\bar{\nabla}_\nu -\delta ^\mu_{\ \nu} \bar{\square}  )\Big) ( \bar{ \square}-r_1 ) 
-2( \pd_\nu\bar{ R}\pd^\mu + \pd^\mu\bar{ R}\pd_\nu)  +2\delta^\mu_{\ \nu}( \pd_\sigma\bar{ R}\pd^\sigma +\bar{ \square}\bar{ R}) \Bigg]\Zc (\bar{\square}) \zeta= 0 \\
\delta E = & \Big[ \partial^\mu\bar{ R}\partial_\mu  +   2 (r_{1} \bar{R} + r_{2}) + 3( \bar{ \square}-r_1 )^2+ (  \bar{ R}+3r_1 )( \bar{ \square}-r_1 )\Big] \Zc (  \bar{ \square} )  \zeta =  0\,. 
\end{aligned}
\label{treq}
\end{equation}
\end{widetext}
where
\begin{equation} \label{zeta}
\zeta  = \delta (\square) \bar{ R} + (\bar{\square} - r_1) \delta R
\end{equation}
and
\begin{equation} \label{z1z2}
\Zc(\bar{\square}) =\frac{\Fc(  \bar{\square})  }{(  \bar{\square} -r_{1})^2}.
\end{equation}
Note that the perturbed EOM Eq.(\ref{treq}) has only one perturbation variable $\zeta$, and we will shortly see that this leads to interesting physics. One may wonder about the absence in Eq.\eqref{treq} of the perturbation of Einstein tensor from GR. This can be easily understood since the following coefficient of Einstein tensor in Eq.\eqref{EoM} vanishes upon using the conditions in Eq.\eqref{condi}
 \begin{equation}
 [M_{p}^{2}+4\Fc(\bar{\square}) \bar{R} ] = 0\,.
 \label{lEqn}
 \end{equation}
From the $i \neq j$ perturbed EOM in Eq.\eqref{treq}, we ultimately obtain the simple EOM for $\zeta$
\begin{equation}
\Zc (\bar{\square}) \zeta =0 \,.
\label{ijeq}
\end{equation}

To deduce  the kinetic structure of $\zeta$ and, therefore, the (background dependent) ghost-free form factor $\Fc (\bar{\square})$,  we compute the second variation of the action Eq. (\ref{action1}) around Eq. (\ref{ansatz}), imposing further the conditions  $\Fc_{1}=0,\,\Fc_2=-\frac{M_p^2}{4}$ 
\begin{widetext}
\begin{equation} \label{qac}
\begin{aligned}
\delta^{2}S= & \int d^{4}x \sqrt{-\bar{g}} \Bigg\{ \frac{M_{p}^{2}}{4} ( \delta_{GR} - \delta^{(2)}R)
-\Lambda \left( \frac{h^{2}}{8} - \frac{1}{4} h_{\mu \nu} h^{\mu \nu} \right)
+\frac{h}{2}\bar{R} \mathcal{F} (\bar{\square})  \delta R
+\frac{h}{2} \bar{R} \delta \cF(\square ) \bar{R}
+\delta R\delta\mathcal{F}(\square)\bar{ R}\\
&\qquad \qquad \qquad \qquad \qquad \qquad \qquad +\bar{R}\delta\mathcal{F}(\square)\delta R
+\bar{ R}\delta^{(2)}\mathcal{F}(\square)\bar{ R}
\Bigg\}
\end{aligned}
\end{equation}
\end{widetext}
where $\sqrt{-\bar{g}} \ \delta_{GR} \equiv \delta^{(2)}(\sqrt{-g} R)$ and $\delta^{(2)}R$ is the second variation of $R$. 
%
%
%
After considerable manipulation of terms in Eq.(\ref{qac}) (see Appendix (\ref{appendix1}) for some useful formulas), the quadratic action remarkably simplifies to
\begin{equation}
\delta^{2} S = \int d^4x \sqrt{-\bar{g}} \ \zeta \Zc (\bar{\square}) \zeta.
\label{s2zet}
\end{equation}
As expected, at the quadratic level of the action, the presence of the sole variable, $\zeta$, is consistent with our earlier result obtained from the linearized EOM in Eq.(\ref{ijeq})\footnote{Our result Eq.(\ref{s2zet}) turns out to be the same as the non-local part of the quadratic action around an inflationary background satisfying $\bar{\square} \bar{R}=r_1 \bar{R}$ \cite{Koshelev:2017tvv}} - by varying $\delta^{2} S$ with respect to $\zeta$, we exactly recover the linearized EOM for $\zeta$.

To clearly expose the physical degrees of freedom in our theory at the level of the quadratic action, we study SVT modes in the context of (1+3) cosmological perturbation theory \cite{Mukhanov:1990me}. We fix the background $\bar{g}_{\mu \nu}$ to be spatially flat Friedmann-Lema\^itre-Robertson-Walker (FLRW) given by the line element $ds^2= -dt^2+a(t)^2d\vec{x}^2$, $a(t)$ being the scale factor, satisfying the ansatz Eq.(\ref{ansatz}). We write the full metric as $g_{\mu \nu} = \bar{g}_{\mu \nu} + h_{\mu \nu}$, where $h_{\mu \nu}$ is the perturbation. In longitudinal gauge, $h_{\mu \nu}$ has components
   \begin{equation} \label{svt}
   h_{00} = a^{2} \left( -2 \phi \right),\
   h_{0i} = a^{2} (\hat{B}_{i} ),\
   h_{ij} = a^{2} (- 2 \psi  \delta_{ij}  + 2 \hat{h}_{ij} ),
   \end{equation}
so that $h_{\mu \nu}$ has $6$ degrees of freedom (2S+2V+2T) after gauge fixing\footnote{The longitudinal gauge removes $2$ scalar and $2$ vector degrees of freedom from the original total of $10$ in $h_{\mu \nu}$} \cite{Craps:2014wga}. Note that $\partial^{i} \hat{B}_{i} = \partial^{i} \hat{h}_{ij} = \hat{h}_{i}^{\ i} = 0$. Given this decomposition, it is straightforward to see that $\zeta$ (defined in Eq.(\ref{zeta})) is a linear combination of $\phi$ and $\psi$, and therefore is effectively a scalar under $SO(3)$. The quadratic action Eq.(\ref{s2zet}) then solely has one scalar propagating degree of freedom $\zeta$. It is easy to understand why the vector $\hat{B}_i$ and tensor mode $\hat{h}_{ij}$ are absent. The background conditions we imposed, see Eq.(\ref{condi}), lead to the removal of vector and tensor modes that could potentially come from the term proportional to $M_p^2$ in Eq.(\ref{qac}). All other terms in $\delta^{2}S$ generate only scalars in terms of $\phi$ and $\psi$. This confirms the result we had obtained earlier from the linearized EOM.

For the theory to have no extra degrees of freedom or ghosts at the quadratic level of the action in Eq.(\ref{s2zet}), the kinetic operator $\Zc( \bar{\square} )$ can have at most a single zero according to the Weierstrass product theorem. For the case of $\Lambda>0$ (see Eq.(\ref{pcc})), we therefore make the following minimal choice for $\Fc (\bar{\square})$ (see Eq.(\ref{z1z2})) satisfying all the necessary conditions in Eq.\eqref{condi}
\begin{equation}
\ba \label{formf}
\Fc (\bar{\square})  =  \frac{1}{M_{s}^{6}} \LF\bar{\square} - m^2\RF\LF \bar{\square} - r_1\RF^2 e^{\gamma (\bar{\square})}
\ea
\end{equation}
where $m^{2}\geq0$ is some arbitrary mass scale and $\gamma$ is an arbitrary entire function of $\bar{\square}/M_{s}^{2}$. This choice ensures that the kinetic term for $\zeta$ has only one zero at $\bar{\square} = m^{2}$. Note that the kinetic term for $\zeta$ has the correct sign to avoid ghosts.
%
Similarly, for the case of $\Lambda<0$, the minimal choice for $\Fc (\bar{\square})$ would be one without any zeros for $\Zc( \bar{\square} )$
\begin{equation}
\ba \label{formf2}
\Fc (\bar{\square})  =  \frac{1}{M_{s}^{4}} ( \bar{\square} - r_1)^2 e^{\gamma (\bar{\square})}.
\ea
\end{equation}
In this case, $\zeta$ contains no poles in the propagator and acts like a $p$-adic scalar \cite{Barnaby:2007ve}.
%
%
%
%
%
%

We now specialize to known bouncing solutions \cite{Biswas:2005qr,Biswas:2010zk,Koshelev:2014voa}. They all satisfy the ansatz Eq.(\ref{ansatz}), are vacuum solutions of the non-local theory Eq.(\ref{action1}), and possess only a single ghost-free scalar mode (and no vector or tensor modes) around the bouncing background at the linearized level when the non-local form factor $\Fc (\bar{\square})$ is chosen appropriately as just discussed
\begin{enumerate}
\item Cosine hyperbolic bounce with $a(t)= a_0 \cosh(\sqrt{{r_1}/{2}}t)$.
In this case, $\Lambda >0$.
\item Exponential bounce with $a(t)= a_0e^{\frac{\lambda}{2}t^2}$. In this case, $\Lambda>0$ for $\lambda>0$ and $\Lambda<0$ for $\lambda<0$. For $\lambda>0$ we have contraction followed by expansion like in a usual bounce. For $\lambda<0$, the universe reaches a maximum scale factor at $t=0$ and then starts to decrease in size. This is also a possible way the Big Bang singularity could be resolved.
\item Cyclic universe: This scenario was numerically studied in \cite{Biswas:2010zk}. It was found that in the case of $\Lambda<0$, there exists a cyclic scenario where the universe undergoes multiple bounces. 
\end{enumerate}

\begin{figure}[h!]
	\includegraphics[scale=0.5]{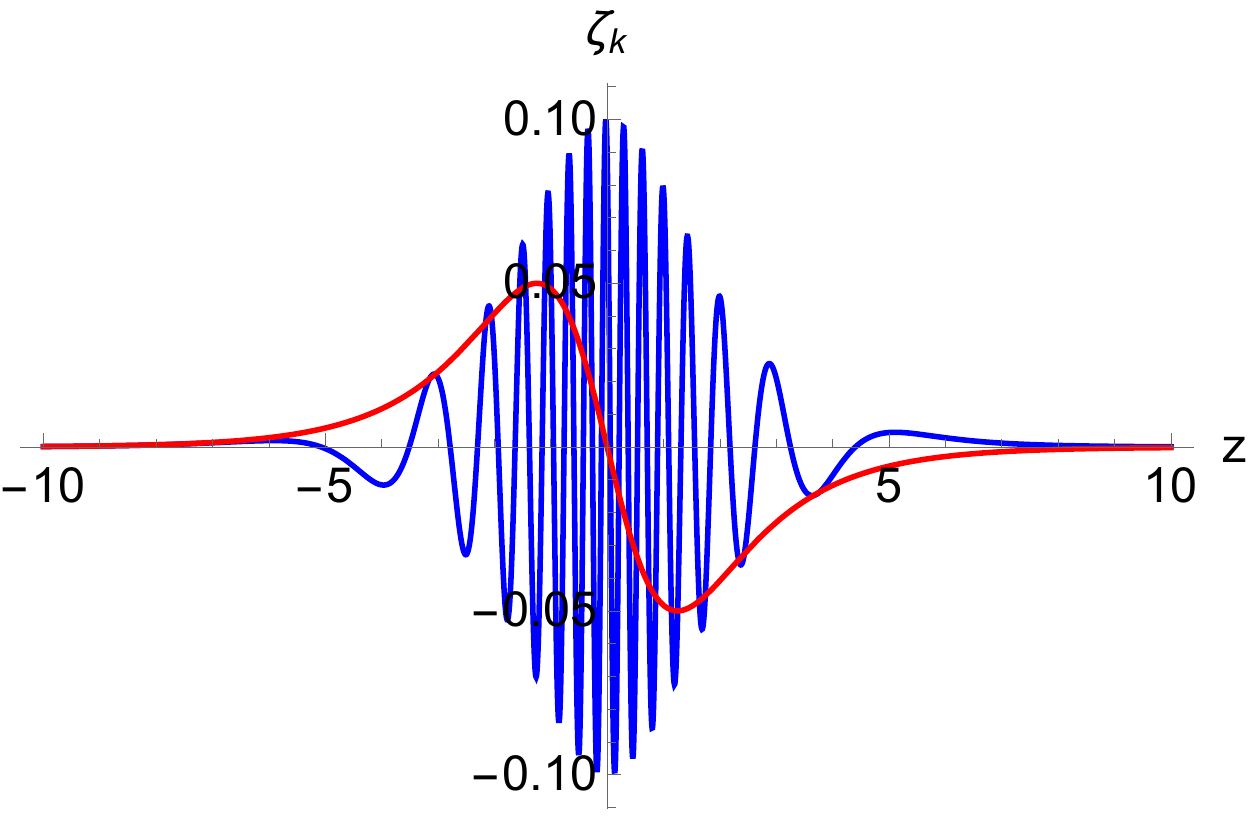}
	\caption{Evolution in cosmic time of a Fourier component of the scalar mode $\zeta_{k}$ for cosine hyperbolic bounce. We have set $a_{0} = 1, r_{1} = m^{2} = M_{s}^{2}, c_1 = c_2 = 0.1M_s$ and $k = 20 M_{s}$ (blue curve) and $k=0$ (red curve).}
	\label{Fig1}
\end{figure}

Let us restrict to the case of $\Lambda>0$ and a cosine hyperbolic bounce, in which case $\Fc (\bar{\square})$ is given by Eq.(\ref{formf}). For a stable bounce, we would require the perturbation $\zeta$ to be well behaved in time. All solutions of the non-local EOM for  $\zeta$ in Eq.(\ref{ijeq}), for $\Fc (\bar{\square})$ given in Eq.(\ref{formf}), are captured by solutions of the local equation
\be \label{localeom}
\LF\bar{\square} - m^2\RF \zeta = 0
\ee
because the non-local factor $e^{\gamma(\bar{\square})}$ does not introduce any new poles in the propagator \cite{Biswas:2005qr,Barnaby:2007ve}. Setting $a_0=1, r_1 = m^2 = M_s^2$ for the cosine hyperbolic bounce, the generic solution of Eq.(\ref{localeom}) in Fourier space looks like 
\begin{equation}
\zeta_{k}(z) =   c_{1} H_{k}^{(+)}(z) + c_2 H_{k}^{(-) }(z)\,,
\label{solzet}
\end{equation}
where
\begin{equation}
\begin{aligned}
H_{k}^{\pm} = &  \frac{\sqrt{e^{\sqrt{2} z}}}{1+e^{\sqrt{2} z}}\left[\frac{\LF 1\pm i\sqrt{e^{\sqrt{2} z}}\RF ^2}{1+e^{\sqrt{2} z}}\right]^{ \sqrt{\frac{ 2k^2}{M_s^2}+1}},
\end{aligned}
\end{equation}
where $z= M_s t$ and $c_{1,2}$ are constants. We can easily deduce that $\zeta_k$ is real for $c_1= c_2$. 
Here, we do not quantize $\zeta$ and merely wish to observe how it evolves classically through the bouncing phase.
We have shown the solution $\zeta_{k}$ in Figure.~\ref{Fig1}, where we can see that it is oscillatory and bounded for the chosen values of $c_{1,2}$ and $k$. This behaviour persists to hold for any value of  $k$ with increasing number of bounded oscillations as $k\to \infty$. 
We do not present here the solutions for $\zeta$ around other bouncing backgrounds mentioned earlier since the nature of perturbations remains very similar to that of cosine hyperbolic bounce.

For the case of cosine hyperbolic bounce, we will have a de Sitter (inflationary) evolution as a late time attractor \cite{Biswas:2005qr,Koivisto:2008xfa,Biswas:2011qe,Biswas:2012bp}. We can observe from Fig.~\ref{Fig1}  that $\zeta\to 0$ in the late time leading to $\LF \bar{\square}-r_1 \RF \delta R \to 0$. This observation confirms that we exactly recover the late time behavior of scalar fluctuations found in Ref.\cite{Biswas:2012bp}.
  
There is also the intriguing possibility of a phase of super inflation after  cosine hyperbolic bounce, and before cosmic inflation. This makes this type of bounce an interesting framework for explaining low multipoles in the observed data for cosmic microwave background radiation \cite{Biswas:2013dry}.  It is important to note that the late time de Sitter stage is significantly different in our case compared to \cite{Biswas:2012bp}, because of the absence of any tensor modes due to our background conditions in Eq.(\ref{condi}). It is worth mentioning here an analogous study of special (A)dS backgrounds with only a scalar propagating degree of freedom, albeit with a different choice of ghost-free form factor found in Ref.\cite{SravanKumar:2019eqt}. 

To summarize, we have found a simple model of non-singular, homogeneous and isotropic bouncing cosmology in vacuum with spatially flat geometry. Moreover, for the background satisfying certain conditions, see Eq.(\ref{condi}), there exists only a scalar perturbation at the linearized level. There are no vector or tensor modes. Finding a ghost-free form factor $\Fc (\bar{\square})$ around a dynamical background that leads to only a scalar propagating mode is non-trivial and the main highlight of this paper. To our knowledge, this is the first such study in infinite derivative theories of gravity. In a particular bouncing scenario, we have explicitly shown the lone scalar propagating degree of freedom is stable and its evolution is oscillatory and bounded.
Of course, these results are valid at the linearized level in perturbation theory, and it would be interesting to see what happens at higher orders.

{\it Acknowledgements:-}
S.K. and A.M. are supported by Netherlands Organization for Scientific Research (NWO) grant no. 680-91-119. J.P. is supported by the China Scholarship Council. We thank R. H. Brandenberger and A. S. Koshelev for useful discussions. Some computations were checked using \href{http://xact.es/}{\textit{xAct}} and \href{http://www.xact.es/xPand/}{\textit{xPand}}\cite{Brizuela:2008ra,Pitrou:2013hga}. S.M. thanks O. Umeh, T. B\"ackdahl and C. Pitrou for their help with \textit{xAct}.

\appendix
\section{}\label{appendix1}
Simplifying the quadratic action Eq.\eqref{qac} and the linearized EOM Eq.\eqref{treq} involves using the formulas
\begin{equation} \label{deltaf1}
\delta \cF (\square ) = \sum_{n=1}^{\infty}  \sum_{l=0}^{n-1} f_{n} \square_{s}^{l} \delta (\square_{s}) \square_{s}^{n-l-1}
\end{equation}
and
\begin{equation}
\ba \label{delta2f1}
\delta^{(2)} \cF (\square ) &= \sum_{n=1}^{\infty}  \sum_{l=0}^{n-1} f_{n} \square_{s}^{l} \delta^{(2)} (\square_{s}) \square_{s}^{n-l-1} \\
&+ 2 \sum_{n=2}^{\infty} \sum_{l=0}^{n-2} \sum_{k=0}^{n-l-2} f_{n} \square_{s}^{l} \delta (\square_{s}) \square_{s}^{k} \delta (\square_{s}) \square_{s}^{n-k-l-2}.
\ea
\end{equation}
Using these, the background ansatz Eq.(\ref{ansatz}) and conditions in Eq.(\ref{condi}), we can simplify the last five terms in the quadratic action by using the following relations
\begin{equation}
\delta( \Fc( \square ) R ) = ( \bar{ \square} -r_1 ) \Zc( \bar{\square} ) \zeta
\label{A3}
\end{equation}
\begin{widetext}
\begin{equation} 
\begin{aligned}
\int d^{4}x\sqrt{-\bar{g}} \ \bar{R}\delta\mathcal{F} (\square)\delta R 
= & \int d^{4}x\sqrt{-\bar{g}} \Bigg[ \Bigg(\bar{R} +\frac{r_{2}}{r_{1}}\Bigg) \delta(\square)\mathcal{Z}(\bar{\square})( \bar{\square}-r_1 )\delta R+  \frac{h}{2} \frac{r_{2}}{r_{1}} ( \cF(\bar{\square}) - f_{0} ) \delta R \Bigg]\,.\\
\int d^{4}x\sqrt{-\bar{g}} \ \bar{R}\delta^{(2)}\mathcal{F} (\square) \bar{ R} 
= & \int d^{4}x\sqrt{-\bar{g}} \Bigg[   \left(\bar{R}+\frac{r_{2}}{r_{1}}\right)\delta(\square) \mathcal{Z}(\bar{\square})
\delta (\square) \bar{R}  \\ &+    \frac{h}{2} \frac{r_{2}}{r_{1}} ( \bar{\square}-r_1 )  \Zc (\bar{\square})   \delta(\square) \bar{ R}
+
\frac{M_{p}^{2}}{4 r_{1}} \frac{h}{2}   \left(\delta(\square)  \bar{R} + \delta^{(2)}(\square) \bar{ R} \right) \Bigg]\,.
\end{aligned}
\label{A4}
\end{equation}
Substituting Eq.(\ref{A3})  and Eq.(\ref{A4}) in Eq.(\ref{qac}), using the following identity for some scalars $X$ and $Y$, performing numerous integration by parts and omitting total derivatives, we finally obtain the result in Eq.(\ref{s2zet})
\begin{equation}
\ba \label{rdeltax}
&\int d^{4} x \sqrt{-\bar{g}} \ Y\delta(\square) X = \int d^{4} x \sqrt{-\bar{g}} \  \left[ X \delta(\square) Y  - \frac{h}{2} \left( Y \bar{\square} X - X \bar{\square} Y\right)  \right].
\ea
\end{equation}
\end{widetext}

%
%
%

\bibliography{ssa}
\bibliographystyle{h-physrev}

\end{document}